\documentclass[prl,reprint,nofootinbib,superscriptaddress]{revtex4-1}
\usepackage{amsmath,amssymb,amsfonts}
\usepackage[dvips]{graphicx}
\usepackage{epsfig}
\usepackage{dcolumn}
\usepackage{bm}
\usepackage{bbm}
\usepackage{xcolor}
\usepackage{lineno}

\usepackage{ulem}

\newcommand{\ket}[1]{|{#1}\rangle}
\newcommand{\bra}[1]{\langle{#1}|}
\newcommand{\inp}[2]{\langle{#1}|{#2}\rangle}

\def\gtap{\ \raise.3ex\hbox{$>$\kern-.75em\lower1ex\hbox{$\sim$}}\ }
\def\ltap{\ \raise.3ex\hbox{$<$\kern-.75em\lower1ex\hbox{$\sim$}}\ }

\begin{document}


\title{Low-energy {\boldmath$\eta$}-nucleon interaction studied with {\boldmath$\eta$} photoproduction off the deuteron}
\author{S. X. Nakamura}
\affiliation{Department of Physics, Osaka University, Toyonaka, Osaka 560-0043, Japan}
\affiliation{
Laborat\'orio de F\'isica Te\'orica e Computacional-LFTC, 
Universidade Cruzeiro do Sul, S\~ao Paulo, SP 01506-000, Brazil
}
\author{H. Kamano}
\affiliation{KEK Theory Center, Institute of Particle and Nuclear Studies (IPNS),
High Energy Accelerator Research Organization (KEK), Tsukuba, Ibaraki
305-0801, Japan}
\affiliation{J-PARC Branch, KEK Theory Center, IPNS, KEK, Tokai, Ibaraki 319-1106, Japan}
\author{T. Ishikawa}
\affiliation{Research Center for Electron Photon Science (ELPH), 
Tohoku University, Sendai, Miyagi 982-0826, Japan}

\begin{abstract}
We develop a reaction model for 
$\eta$ photoproduction off the deuteron ($\gamma d\to\eta pn$), and study
the reaction
at a special kinematics, where 
the photon beam energy is $\sim 0.94$~GeV 
and the scattered proton is detected at $\sim 0^\circ$,
for the purpose of determining 
the $\eta$-nucleon scattering
length ($a_{\eta N}$) and effective range ($r_{\eta N}$).
In this kinematics, 
the $\eta$-nucleon elastic rescattering is significantly enhanced while
other background mechanisms are suppressed.
We show that a ratio $R$, the $\gamma d\to\eta pn$ cross section
divided by the $\gamma p\to\eta p$ cross section convoluted with the
proton momentum distribution in the deuteron, 
has a very good resolving power of 
$a_{\eta N}$ and $r_{\eta N}$.
We conclude that 
the $R$ data with 5\% error, binned in 1 MeV width of the $\eta$-neutron invariant
mass,
can determine
${\rm Re}[a_{\eta N}]$ (${\rm Re}[r_{\eta N}]$) at the
precision of $\sim\pm$0.1~fm ($\sim\pm$0.5~fm),
significantly narrowing down the previously estimated ranges of the parameters.
To arrive at the conclusion, it is essential to
use the $\gamma d\to\eta pn$ reaction model equipped with
elementary amplitudes 
that are well constrained by $\pi N$ and $\gamma N$ reaction data
through a sophisticated coupled-channels analysis.
This result strongly motivates 
the Research Center for Electron Photon Science (ELPH)
at Tohoku University 
to measure $R$.
\end{abstract}


\maketitle

The low-energy interaction between the $\eta$ meson and the nucleon ($N$) 
is, as with the $\pi N$ interaction, a basic feature of the meson-baryon dynamics.
It is characterized by the two
complex parameters,
the scattering length $a_{\eta N}$ and effective range $r_{\eta N}$,
defined through an effective-range expansion of the $S$-wave $\eta N$
scattering amplitude:
$F_{\eta N}(k) = [k\cot\delta(k) - ik]^{-1}$
with $k\cot\delta(k) = a_{\eta N}^{-1} +(1/2)r_{\eta N}k^2  + O(k^4)$,
where $k$ is the on-shell $\eta$ momentum in the center-of-mass (c.m.) frame
and $\delta(k)$ the phase shift.
Because $a_{\eta N}$ determines the attractive or repulsive nature of 
the $\eta N$ interaction at $k\sim 0$, the 
existence of 
exotic
$\eta$-mesic nuclei, which have been actively searched for experimentally,
hinges on its value~\cite{etan8,etan9}.
Accurate values of 
$a_{\eta N}$ and $r_{\eta N}$ can also greatly help determine 
the pole position of the $S$-wave $N(1535)1/2^-$ resonance,
the first spin-$1/2$
negative-parity excitation of the nucleon; the pole is known to be near
the $\eta N$ threshold but 
its accurate position is still uncertain~\cite{pdg}.
It is known that
the $S$-wave scattering parameters can well determine
an $S$-wave resonance pole near threshold~\cite{pis,knlskp2}.

Despite its important role
in nuclear and hadron physics,
the low-energy $\eta N$ interaction has not been well understood yet.
This is attributed mainly to the fact that neither direct $\eta N$ scattering 
experiments nor $X$-ray measurements from $\eta$-mesic atoms 
are possible due to the neutral and unstable nature of $\eta$,
and thus one has to rely on indirect information.
Previous works have attempted to extract $a_{\eta N}$ and $r_{\eta N}$ 
by analyzing the $\pi N\to \pi N, \eta N$ and $\gamma N\to \pi N, \eta N$
reaction data
that have a sensitivity to the $\eta N$ interaction 
through coupled-channels effects~\cite{etan9}. 
The $pn \to \eta d$ reaction has also been analyzed to extract 
the $\eta N$ interaction embedded in the strongly interacting
$\eta NN$ system~\cite{etan5}.
These analyses gave fairly convergent results for 
the imaginary parts of $a_{\eta N}$ and $r_{\eta N}$,
the values of which fall into 
${\rm Im} [a_{\eta N}]=0.2$--0.3~fm 
and 
${\rm Im} [r_{\eta N}]=-1$--0~fm, 
respectively~\cite{etan9,gw-2005}.
However, their real parts scatter in a rather wide range:
${\rm Re} [a_{\eta N}]=0.2$--0.9~fm and
${\rm Re} [r_{\eta N}]=-6$ to +1~fm.
The large model-dependence
of the previously extracted ${\rm Re} [a_{\eta N}]$ and ${\rm Re} [r_{\eta N}]$
originates from the difficulty of isolating the $\eta N$ scattering amplitudes 
from other mechanisms involved in the reactions analyzed.
Therefore, 
it is highly desirable to utilize reactions 
in which mechanisms associated with the $\eta N$ elastic rescattering 
are significantly enhanced while other background mechanisms
being suppressed.

\begin{figure*}
\includegraphics[clip,width=0.98\textwidth]{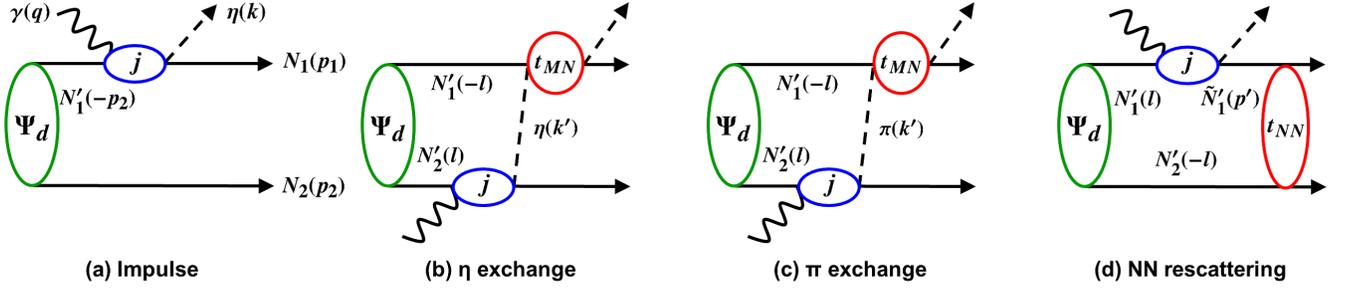}
\caption{\label{fig:diag}
Diagrammatic representation of reaction mechanisms considered in this work for $\gamma d\to\eta N_1 N_2$:
(a) impulse, (b) $\eta$-exchange, (c) $\pi$-exchange, and
(d) $NN$ rescattering mechanisms.
Labels for particles along with their momenta
in the Lab frame are indicated.
The external lines are the same for all the diagrams and thus their
 labels are indicated in (a) only.
Also, $\bm{k'} = \bm{q}-\bm{p_2}+\bm{l}$ and $\bm{p'} = \bm{q}-\bm{k}+\bm{l}$.
}
\end{figure*}

To meet this demand,
an $\eta$ photoproduction experiment~\cite{plan} is planned 
at the Research Center
for Electron Photon Science (ELPH), Tohoku University.
In this experiment, 
a deuteron target is irradiated with
a photon beam at the laboratory energy of $E_\gamma\sim 0.94$~GeV~\cite{tagger,bpm},
and the recoil proton from the $\gamma d\to \eta pn$ reaction
is detected at 
$\theta_p\sim 0^\circ$
 from the photon direction.
At this chosen kinematics, 
an $\eta$ is likely to be produced almost at rest,
being expected to strongly interact with the spectator neutron.
Meanwhile, the struck proton goes away with a large momentum, leaving
little chance to interact with the $\eta$ and neutron.
This seems an ideal kinematical condition, 
to which we refer as the ELPH kinematics,
to determine the low-energy $\eta N$ scattering parameters.
In this Rapid Communication, we show with a theoretical analysis that
a combined cross-section data for $\gamma d\to \eta pn$ and $\gamma p\to \eta p$
expected to be taken at the ELPH experiment
would indeed lead to significant reduction of the uncertainty 
of $a_{\eta N}$ and $r_{\eta N}$ previously extracted,
thereby providing
crucial constraints on the existence of $\eta$-mesic nuclei
and the properties of $N(1535)1/2^-$.

The possibility of extracting $a_{\eta N}$ from 
$\gamma d\to \eta pn$ data was first explored by Sibirtsev
{\it et al.}~\cite{juelich2002a}, and
a fairly large $a_{\eta N}$-dependence of
$\eta$ angular and momentum distributions was shown.
However, a subsequent work by Fix {\it et al.}~\cite{mainz2004} 
found a significantly less pronounced $a_{\eta N}$-dependence than 
those of Ref.~\cite{juelich2002a}, leading to the conclusion 
that it is practically impossible to extract $a_{\eta N}$ from 
$\gamma d\to \eta pn$ 
data.
Thus, until the present work, 
no practically useful connection has been made
between $\gamma d\to \eta pn$ data and $a_{\eta N}$.
We note that these pioneering works~\cite{juelich2002a,mainz2004}
 examined $\gamma d\to \eta pn$ near the threshold
($E_\gamma<$ 0.7~GeV)
while we study the reaction in 
rather different kinematics
($E_\gamma\sim 0.94$~GeV; $\theta_p\sim 0^\circ$).

We study $\gamma d \to \eta p n$ relevant to the ELPH experiment
with a model based on the impulse and the first-order rescattering
mechanisms 
as 
depicted in Fig.~\ref{fig:diag}.
The $\eta$-exchange mechanism [Fig.~\ref{fig:diag}(b)]
contains the $\eta N \to \eta N$ subprocess we are interested in,
while the other mechanisms (the impulse [Fig.~\ref{fig:diag}(a)], 
$\pi$-exchange [Fig.~\ref{fig:diag}(c)],
and $NN$-rescattering [Fig.~\ref{fig:diag}(d)] mechanisms)
are backgrounds for our purpose.
With the momenta defined in Fig.~\ref{fig:diag},
the amplitudes 
for
$T_{\rm imp}$ (impulse),
$T_\eta$ ($\eta$-exchange),
$T_\pi$ ($\pi$-exchange), and
$T_N$ ($NN$-rescattering),
are explicitly written in the laboratory frame as
\begin{widetext}
\begin{equation}
\begin{split}
T_{{\rm imp}} =
&
\sqrt{2}
\sum_{s_1'}
\bra{\eta(\bm{k})\, N_1(\bm{p}_1,s_1,t_1)} 
j(M_{\eta N_1})
\ket{\gamma(\bm{q},\mu)\, N_1'(-\bm{p}_2 ,s_1',t_1)}
\inp{N_1'(-\bm{p}_2,s_1',t_1)\, N_2 (\bm{p}_2,s_2,t_2)}{\Psi_d(s_d)}
\ ,
\end{split}
\label{eq:amp_imp}
\end{equation}
\begin{equation}
\begin{split}
T_{M(=\eta,\pi^\pm,\pi^0)} =
&\sqrt{2}
\sum_{s_1',s_2'}
\sum_{t_1',t_2'}
\int d\bm{l}
\bra{\eta(\bm{k})\, N_1(\bm{p}_1,s_1,t_1)}
t_{MN}(M_{\eta N_1})
\ket{M(\bm{q}-\bm{p}_2+\bm{l})\, N_1'(-\bm{l},s'_1,t_1')}
\\
&\times
{\bra{M(\bm{q}-\bm{p}_2+\bm{l})\, N_2(\bm{p}_2,s_2,t_2)} 
j(W)
\ket{\gamma(\bm{q},\mu)\, N_2'(\bm{l},s_2',t_2')}
\over 
E-E_N(\bm{p}_2)-E_N(-\bm{l})-E_{M}(\bm{q}-\bm{p}_2+\bm{l})+i\epsilon}
\inp{N_1'(-\bm{l},s_1',t_1')\, N_2' (\bm{l},s_2',t_2')}{\Psi_d(s_d)}
\ ,
\label{eq:amp_MN}
\end{split}
\end{equation}
\begin{equation}
\begin{split}
T_{N} =
&\sqrt{2}
\sum_{s_1',\tilde s_1',s_2'}
\int d\bm{l}\
\bra{N_1(\bm{p}_1,s_1,t_1)\, N_2(\bm{p}_2,s_2,t_2)}
t_{NN}(M_{N_1N_2}) 
\ket{\tilde N'_1(\bm{q}-\bm{k}+\bm{l},\tilde s'_1,t_1)\, N'_2(-\bm{l},s'_2,t_2)}
\\
&\times
{\bra{\eta(\bm{k})\, \tilde N_1'(\bm{q}-\bm{k}+\bm{l},\tilde s_1',t_1)} 
j(W)
\ket{\gamma(\bm{q},\mu)\, N_1'(\bm{l},s_1',t_1)}
\over E-E_N(\bm{q}-\bm{k}+\bm{l})-E_N(-\bm{l})-E_\eta(\bm{k})+i\epsilon}
\inp{N_1'(\bm{l},s_1',t_1)\, N_2' (-\bm{l},s_2',t_2)}{\Psi_d(s_d)}
\ ,
\label{eq:amp_NN}
\end{split}
\end{equation}
\end{widetext}
plus the exchange terms obtained from
Eqs.~(\ref{eq:amp_imp})--(\ref{eq:amp_NN}) 
by flipping the overall sign 
and interchanging 
all subscripts 1 and 2 such as 
$\{N_1^{(\prime)},\bm{p}_1,s^{(\prime)}_1,t^{(\prime)}_1\}\leftrightarrow \{N_2^{(\prime)},\bm{p}_2,s^{(\prime)}_2,t^{(\prime)}_2\}$.
The elementary (off-shell) amplitudes for
photoproduction, 
meson-baryon, and $NN$ rescattering are denoted by 
$\bra{M N} j \ket{\gamma N'}$,
$\bra{MN}t_{MN}\ket{M'N'}$,
and $\bra{N_1N_2}t_{NN}\ket{N'_1N'_2}$, 
respectively. 
Here, $\ket{\Psi_d(s_d)}$ is the deuteron state at rest with spin projection $s_d$;
$\ket{N(\bm{p},s,t)}$ the nucleon state with momentum $\bm{p}$ and spin and isospin projections $s$ and $t$;
$\ket{\gamma(\bm{q},\mu)}$ the photon state with momentum $\bm{q}$ and polarization $\mu$;
and $\ket{M(\bm{k})}$ ($M$=$\eta, \pi^\pm,\pi^0$) the pseudoscalar meson state with momentum $\bm{k}$.
The total scattering energy $E$ of the system in the laboratory frame is given by the sum of the
photon laboratory energy and the deuteron rest mass, $E=E_\gamma+m_d$, 
and the invariant masses of the two-body subprocesses in the above equations are defined to be
$M_{\eta N_1}=\left\{[E_\eta (\bm{k})+E_{N}(\bm{p}_1)]^2-(\bm{k}+\bm{p}_1)^2\right\}^{1/2}$,
$W=\left\{[E-E_{N}(-\bm{l})]^2-(\bm{l}+\bm{q})^2\right\}^{1/2}$, and
$M_{N_1 N_2}=\left\{[E_{N} (\bm{p}_1)+E_{N}(\bm{p}_2)]^2-(\bm{p}_1+\bm{p}_2)^2\right\}^{1/2}$,
where $E_x(\bm{p})=\sqrt{m_x^2+\bm{p}^2}$ with $m_x$ being the mass of a particle~$x$.
\begin{figure}
\includegraphics[clip,width=0.48\textwidth]{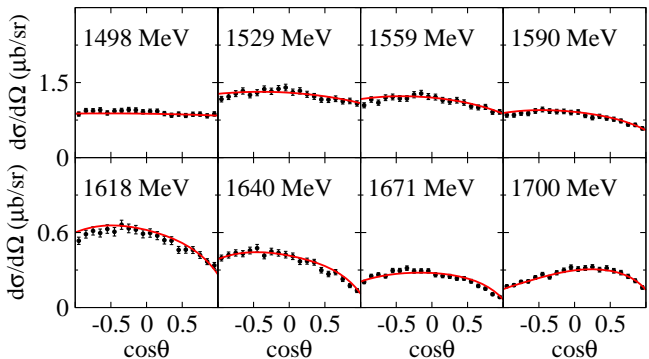}
\caption{\label{fig:gp-etap}
Differential cross sections for $\gamma p\to\eta p$ from the DCC model~\cite{knls16} in
comparison with data~\cite{gp-etap-data}
at selected invariant masses of the $\gamma p$ system.
The corresponding values of the invariant mass are indicated in each panel.
}
\end{figure}

The above definition on $W$ would call for an explanation, 
because other choices of $W$ have also been seen in the literature~\cite{w-choice}. 
We calculate the mechanisms [Fig.~\ref{fig:diag}(a)--\ref{fig:diag}(d)] in a manner
consistent with the well-established Faddeev framework up to the
truncated higher order terms.
The Faddeev framework uniquely specifies
the energy (and thus $W$) of an interacting
two-body subsystem in an intermediate state.
A requirement is to combine the equation with elementary 
(off-shell)
amplitudes
calculated consistently with the Faddeev framework.
Our elementary amplitudes are, as discussed shortly,
calculated with meson-nucleon and nucleon-nucleon potentials
that perfectly fit the Faddeev framework.
Meanwhile, another prescription of $W$ corresponds to another three-dimensional
scattering equation that should work with its own consistent elementary
amplitudes 
but not with ours.
Therefore,
it does not make sense for us to use
the other choices of $W$.
However, if one uses dynamical
inputs that are not consistent with any of the scattering frameworks,
as has been the case in most of the past works,
there is no principle to determine $W$, and thus 
various choices need to be considered.

We now specify our $\gamma d \to \eta p n$ reaction model
to evaluate Eqs.~(\ref{eq:amp_imp})--(\ref{eq:amp_NN}).
The model must be built with reliable amplitudes 
for elementary $\gamma N \to MN$, $MN\to M'N$, and $NN\to NN$ processes
with $M^{(\prime)}$=$\pi,\eta$, 
as well as with a realistic deuteron wave function, 
so that we can 
reliably isolate the amplitude for the $\eta N \to \eta N$ subprocess
from data with 
well-predicted contributions from all the other background mechanisms.
Regarding $\gamma N \to MN$ and $MN \to M'N$ amplitudes,
we employ those generated with a dynamical coupled-channels (DCC)
model~\cite{knls13,knls16}.
The DCC model is a multichannel unitary model for the $\pi N$ and $\gamma N$
reactions in the nucleon resonance region. 
It was constructed fitting $\sim 27,000$ data points, and 
successfully describes~\cite{knls13,knls16,pipin}
$\pi N \to \pi N, \pi\pi N, \eta N, K\Lambda, K\Sigma$ and
$\gamma N \to \pi N, \pi\pi N, \eta N, K\Lambda, K\Sigma$ reactions
over the energy region from the thresholds up to 
$\sqrt{s}\lesssim 2.1$~GeV.
As an example, we present the
$\gamma p\to \eta p$ differential cross sections calculated with the DCC
model of Ref.~\cite{knls16} in Fig.~\ref{fig:gp-etap}.
The figure shows a very good agreement between the model and 
data~\cite{gp-etap-data} over the energy
region relevant to the following calculations of $\gamma d \to \eta p n$.
This verifies that
the most important $\gamma p \to \eta p$ amplitudes out of the
elementary amplitudes for describing $\gamma d \to \eta p n$
are well constrained by the data.
This DCC model predicts the $\eta N$
scattering parameters
to be $a_{\eta N} = 0.75 + 0.26i$~fm and $r_{\eta N} = -1.6 - 0.6i$~fm,
which are consistent with the previously estimated ranges.
As for the deuteron wave function and the $NN$ scattering amplitudes, 
we employ those generated with the CD-Bonn potential~\cite{cdbonn}.

Previous models~\cite{mainz1997,juelich2002a,mainz2004,mainz2015}
also took account of the mechanisms shown in Fig.~\ref{fig:diag};
the $\pi$-exchange mechanism was considered only in Ref.~\cite{mainz2015}.
However, comparing the elementary amplitudes implemented in the previous models,
the DCC model possesses unique and sound features such as:
(i) the model describes {\it all} the meson-baryon and photoproduction amplitudes relevant to
$\gamma d \to \eta p n$ in a unified manner;
(ii) the model generates, by construction, off-shell amplitudes
that are well-suited for working with the Faddeev framework.
We also note that 
a simple $\gamma p \to\eta p$ model including only the
$S_{11}(1535)$-excitation mechanism~\cite{mainz1997,juelich2002a}
is not enough for practically describing $\gamma d\to \eta pn$ at the ELPH kinematics
because the $\gamma p \to\eta p$ amplitudes of $\sqrt{s}=1.6$--1.7~GeV
give a large contribution.

\begin{figure}
 \includegraphics[clip,width=0.49\textwidth]{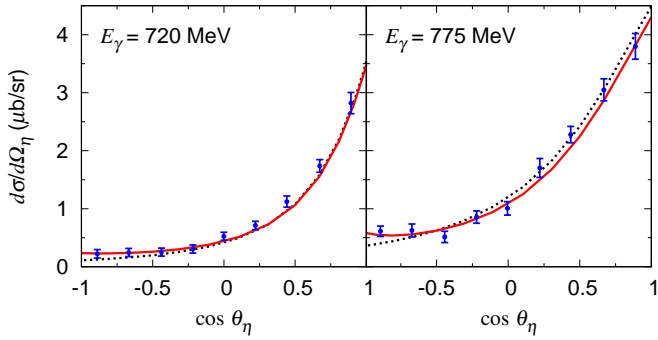}
\caption{\label{fig:eta-data}
Angular distributions of $\eta$ in $\gamma d\to \eta pn$
in the $\gamma d$ c.m. frame.
The photon laboratory energy ($E_\gamma$) is indicated in each panel.
The solid curves are from the full calculation, while
the dotted curves are obtained with the impulse mechanism only. 
The data are for the semi-inclusive $\gamma d\to \eta X$ process~\cite{eta-data};
the coherent contribution is negligible here~\cite{eta_coh}.
}
\end{figure}
The setup described above allows us to make a parameter-free prediction for
the $\gamma d\to\eta pn$ cross sections.
We thus confront our model predictions
with existing data,
thereby assessing the validity of the model.
In Fig.~\ref{fig:eta-data}, we show 
the $\eta$ angular distribution at $E_\gamma = 720$ and 775~MeV
from our DCC-based model 
with and without the rescattering contributions along with the data.
Our parameter-free 
prediction is found to be in an excellent agreement with the data.
A slight enhancement in the backward direction
due to the $\eta N\to \eta N$ rescattering is
important for this agreement. 
Fix {\it et al.}~\cite{mainz2015} also have done 
a comparable calculation, and found a rather minor role of 
the $\eta N \to \eta N$ rescattering mechanism 
in the $\eta$ angular distribution at these energies.
The slight underestimation of their results at backward angles
(Fig.~5 of Ref.~\cite{mainz2015}) is
likely to be ascribable to the different
$\eta N$ scattering lengths;
$a_{\eta N} = 0.75 +  0.26i$~fm in our model and 
$a_{\eta N} = 0.5 +  0.32i$~fm in Ref.~\cite{mainz2015}.
Regarding the cross sections with the impulse mechanism only, 
our result is close to that of Ref.~\cite{mainz2015} while
significantly smaller than that of Ref.~\cite{juelich2002a}.
See Ref.~\cite{mainz2004} for a detailed discussion on 
the difference with Ref.~\cite{juelich2002a}.

\begin{figure}
 \includegraphics[clip,width=0.485\textwidth]{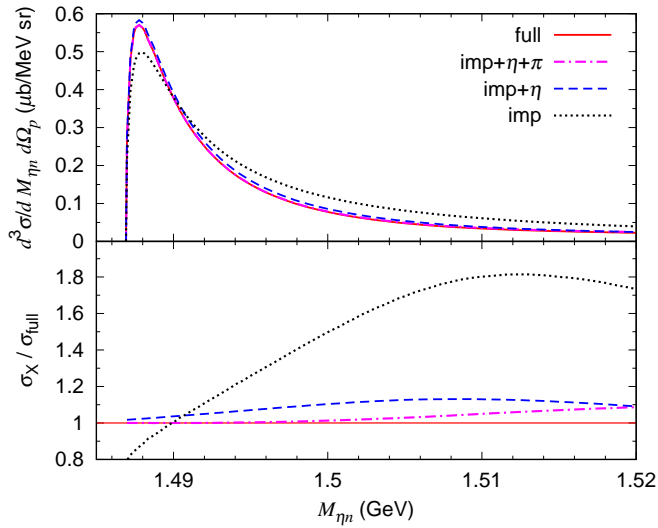}
\caption{\label{fig:eta} 
(Top)
Threefold differential cross section,
$d^3\sigma/dM_{\eta n} d\Omega_p$,
for $\gamma d \to \eta pn$
at $E_\gamma = 0.94$~GeV and $\theta_p=0^\circ$,
plotted as a function of $M_{\eta n}$.
The results are from 
the full calculation (solid curve),
the impulse mechanism only (dotted curve), 
the impulse and $\eta$-exchange mechanisms (dashed curve),
and
the impulse, $\eta$- and $\pi$-exchange mechanisms (dash-dotted curve).
The dash-dotted curve falls almost exactly on the solid curve. 
(Bottom) Ratios of the differential cross sections calculated with the various
mechanisms to those from the full calculation.
}
\end{figure}
Now let us consider the $\gamma d\to\eta pn$ reaction at the ELPH kinematics with
$E_\gamma=0.94$~GeV and $\theta_p=0^\circ$.
In Fig.~\ref{fig:eta}(top),
our model predictions for
the threefold differential cross section, $d^3\sigma/dM_{\eta n} d\Omega_p$,
are presented as a function of $M_{\eta n}$.
We find that the dominant contribution is from the impulse mechanism
[Fig.~\ref{fig:diag}(a)] that contains 
the $\gamma p\to \eta p$ amplitudes, 
while the
$\gamma n\to \eta n$ amplitudes negligibly contribute.
The $\eta$-exchange mechanism [Fig.~\ref{fig:diag}(b)]
has a substantial contribution to the cross section, 
which changes the impulse result by $-$40 to +20\% 
[difference between the dashed and dotted curves in Fig.~\ref{fig:eta}(bottom)].
Meanwhile, the $\pi$-exchange [Fig.~\ref{fig:diag}(c)] 
contribution is smaller, and suppresses the cross sections
by $\ltap$9\%
(difference between the dashed and dash-dotted curves).
The $NN$ rescattering [Fig.~\ref{fig:diag}(d)] contribution
(deviation of the dash-dotted curve from 1)
is very small for $M_{\eta n}\ltap 1.5$~GeV.
This feature is what we expect to find in this special kinematics.
The $\pi$-exchange mechanism is strongly suppressed
even though the elementary 
$\gamma p \to \pi N$ amplitude is significantly larger
than that of $\gamma p \to \eta p$ at the considered energies.
This is the exchanged pions have
rather large momenta near their on-shell, picking up
high-momentum components with very small probabilities in the deuteron wave function. 
The $NN$-rescattering mechanism is hindered by the same kinematical reason, 
and also by the rather weak $NN$ scattering at this kinematics where 
the $NN$ relative momentum is large. 

We have shown that 
the $\gamma d\to \eta pn$ in the ELPH kinematics for
$M_{\eta n}\ltap 1.5$~GeV are described with the impulse and
$\eta$-exchange mechanisms 
and with the smaller 
(almost negligible)
correction from the
$\pi$-exchange ($NN$-rescattering) mechanism.
This indicates that the proton is well separated from interacting with
the $\eta n$ system, and thus
multiple rescatterings beyond the first-order rescattering 
[Figs.~\ref{fig:diag}(b)--\ref{fig:diag}(d)]
should be safely neglected in this kinematical region.
We have also confirmed that 
an off-shell momentum effect associated with the $\eta n\to \eta n$
scattering amplitude is very small and that 
$\eta n\to \eta n$ partial wave amplitudes
higher than the $S$ wave give
negligibly small contributions.
These facts allow us to modify the full $\gamma d \to \eta p n$ model by 
replacing the $\eta n$ scattering amplitude with
the $S$-wave one parametrized with $a_{\eta N}$ and $r_{\eta N}$,
and then to determine these parameters 
through analyzing the forthcoming ELPH data.
To make contact with the ELPH data, we need to take one more
step because 
the data are actually given in a form of the ratio,
denoted by $R_{\rm expt}$,
of the measured cross sections for $\gamma d\to \eta pn$ divided by those
for $\gamma p\to \eta p$ convoluted with the proton momentum distribution
in the deuteron.
This is for removing systematic uncertainties of the acceptance from the detector coverage.
Thus, from the theoretical side, the corresponding quantity to calculate is:
\begin{eqnarray}
\label{eq:Ratio}
R_{\rm th}(M_{\eta n}) = {d^3\sigma_{\rm full}/dM_{\eta n} d\Omega_p |_{\theta_p=0^\circ}
\over d^3 \sigma_{\rm imp}/dM_{\eta n} d\Omega_p|_{\theta_p=0^\circ}} 
\ ,
\end{eqnarray}
where $\sigma_{\rm full}$ ($\sigma_{\rm imp}$) is 
calculated with
the modified full model (the impulse term only).
The remaining questions to address are how sensitively $R_{\rm th}$ changes
as $a_{\eta N}$ and $r_{\eta N}$ are varied, and how well 
$R_{\rm expt}$ with a certain error can determine
$a_{\eta N}$ and $r_{\eta N}$.

\begin{figure}
 \includegraphics[clip,width=0.485\textwidth]{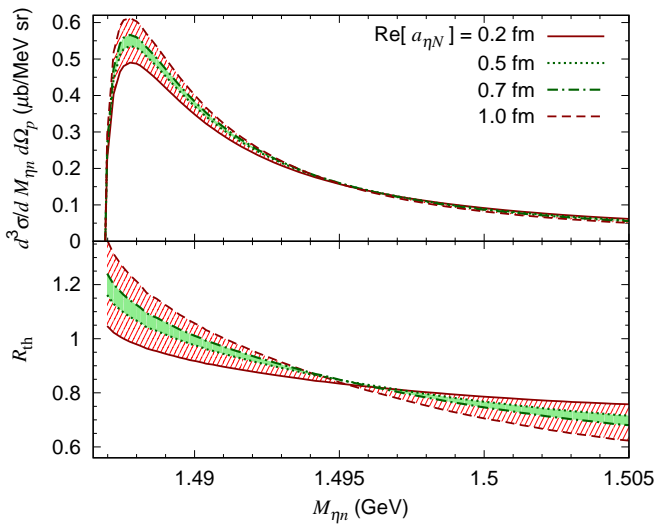}
\caption{\label{fig:eta-a} 
(Top) Re[$a_{\eta n}$]-dependence of $\gamma d\to \eta pn$ differential
 cross sections at $E_\gamma=0.94$~GeV and $\theta_p=0^\circ$
 calculated with the full model.
The curves are obtained with 
${\rm Re}[a_{\eta n}]=0.2$, 0.5, 0.7, and 1.0 fm;
${\rm Im}[a_{\eta n}]=0.25$~fm and $r_{\eta n}=0$.
(Bottom) The quantity $R_{\rm th}$ defined in Eq.~(\ref{eq:Ratio})
for various values of Re[$a_{\eta n}$].
}
\end{figure}
First we vary ${\rm Re}[a_{\eta N}]$ over
0.2 -- 1.0~fm, 
with fixed values of 
${\rm Im}[a_{\eta N}]=0.25$~fm and $r_{\eta N}=0$~fm.
At the ELPH kinematics and $M_{\eta n}\le 1.505$~GeV,
the obtained cross sections are mostly within the red striped region 
shown in Fig.~\ref{fig:eta-a}(top).
The corresponding variation of $R_{\rm th}$ is shown in
Fig.~\ref{fig:eta-a}(bottom) where the 
sensitivity to the variation of ${\rm Re}[a_{\eta N}]$ 
is more clearly seen.
As the striped bands show, 
the cross section and thus $R_{\rm th}$
changes by $\sim$25\% at the quasi-free (QF)
peak position
at $M_{\eta n}\sim 1.488$~GeV.
Meanwhile, the green solid bands,
which are covered when ${\rm Re}[a_{\eta N}]$ is varied by $\pm 0.1$~fm
from 0.6~fm, have the widths of $\sim$5\% at the QF peak. 
The result indicates that 
$R_{\rm expt}$ data of 5\% error per MeV bin,
which is achievable in the planned ELPH experiment~\cite{plan},
can determine ${\rm Re}[a_{\eta N}]$ 
at the precision of $\sim\pm 0.1$~fm, 
significantly narrowing down the current uncertainty.

\begin{figure}
 \includegraphics[clip,width=0.485\textwidth]{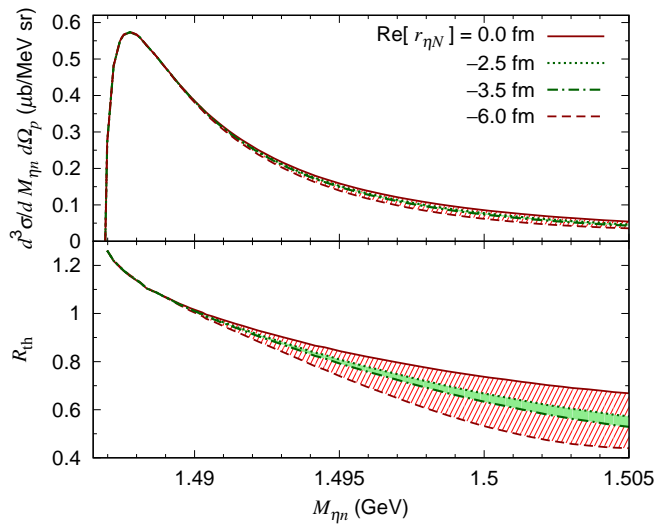}
\caption{\label{fig:eta-r} 
Similar presentation to Fig.~\ref{fig:eta-a}, but using
Re$[r_{\eta n}]=0$~fm (solid), $-$2.5~fm (dotted),  $-$3.5~fm (dash-dotted), 
and $-$6~fm (dashed); 
$a_{\eta n}=0.75+0.26i$~fm
and ${\rm Im}[r_{\eta n}]=0$~fm are fixed.
}
\end{figure}

Next we vary ${\rm Re}[r_{\eta N}]$ over a rather broad range
of the current estimates,
$-$6 -- 0~fm; 
the scattering length is fixed at
$a_{\eta n}=0.75+0.26i$~fm, 
the value from the latest DCC analysis~\cite{knls16};
${\rm Im} [r_{\eta N}]=0$~fm.
The corresponding changes of the cross section and $R_{\rm th}$
cover the red striped region in Fig.~\ref{fig:eta-r}.
Because $r_{\eta N}$ plays no role very close to the $\eta N$ threshold,
its effect
starts to be visible at $\sim$5~MeV above the threshold.
The red striped (green solid) band of $R_{\rm th}$ shows that
$R_{\rm th}$ at $M_{\eta n}=1.5$~GeV 
changes by $\sim$30\% ($\sim$5\%)
when ${\rm Re}[r_{\eta N}]$ is varied over
$-$6 -- 0~fm ($-$3.5 to $-$2.5~fm).
Therefore, 
$R_{\rm expt}$ data of 5\% error per MeV bin can also determine 
${\rm Re}[r_{\eta N}]$ at the precision of $\ltap\pm$0.5~fm, 
significantly improved precision over the current estimates.

Regarding the imaginary part,
we vary ${\rm Im}[a_{\eta N}]$ in the range of
0.2 -- 0.3~fm, the currently estimated range, 
and with ${\rm Re}[a_{\eta N}]=0.6$~fm and $r_{\eta N}=0$~fm.
The cross sections and $R_{\rm th}$ change at most 5\%.
When varying ${\rm Im}[r_{\eta N}]$ over the currently estimated range, 
$-$1 -- 0~fm, 
with $a_{\eta n} = 0.75 + 0.26i$~fm and ${\rm Re}[r_{\eta n}]=0$~fm being fixed,
we found a similar situation.

We argue that theoretical uncertainties
hardly affect the above results.
A major part of the uncertainty of the $\gamma d\to \eta pn$ cross section
is from the elementary $\gamma p\to \eta p$ amplitudes that take over
errors ($\sim\pm$5\%) from $\gamma p\to \eta p$ data fitted.
However,
what we need in analyzing the ELPH data
is not the cross section itself but
$R_{\rm th}$ in which 
theoretical uncertainty in the cross section is largely canceled out.
We have confirmed that $R_{\rm th}$ is very stable ($\ltap$0.1\%) even
when the overall magnitude of the $\gamma p\to \eta p$ amplitudes is varied over $\pm 3$\%.
Another possible source of the uncertainty is 
the subthreshold $\gamma p\to \eta p$ amplitudes which are not
well-constrained by the data. 
However, at the ELPH kinematics,
the cross sections (and thus $R_{\rm th}$)
are found to hardly change
($\ltap$0.1\% at the QF peak; $\ltap$1\% for $M_{\eta n}\le 1.505$~GeV) 
even when the subthreshold contributions are omitted. 
We have also studied 
the model dependence of the deuteron wave function. 
We used those of the CD-Bonn~\cite{cdbonn}, Nijmegen I~\cite{nij}, and
Reid93~\cite{nij} models, and found a rather good
stability ($<$~0.5\% at the QF peak; \ltap 1\% at 
$M_{\eta n}\sim 1.5$~GeV) of $R_{\rm th}$.

Finally, we make clear what we have advanced from the previous
investigations~\cite{juelich2002a,mainz2004} on 
extracting $a_{\eta N}$ from $\gamma d\to\eta pn$ data.
For this purpose, 
it would be illustrative to compare our main result (Fig.~\ref{fig:eta-a}) 
with Fig.~6 (bottom) of Ref.~\cite{mainz2004} that also shows
the $a_{\eta N}$-dependence of
$\gamma d\to \eta pn$ differential cross sections 
at a fixed proton angle.
Despite the similarity, the authors of
Ref.~\cite{mainz2004} were concerned with 
the cross section shape
while we utilize the absolute values of
$R_{\rm th}$ that has a significantly better 
sensitivity to the $\eta N$ scattering parameters.
What enables us to utilize the $R_{\rm th}$ values
is our very well-controlled calculation as follows.
At the kinematics chosen in Ref.~\cite{mainz2004}
($E_\gamma=670$~MeV, $\theta_p=18^\circ$, $M_{\eta n}\sim m_\eta+m_n$),
according to our model, we found:
(i) the elementary $\gamma n \to \eta n$ amplitudes 
give a contribution comparable to that from the 
$\gamma p \to \eta p$ amplitudes; 
(ii) the subthreshold $\gamma p \to \eta p$ amplitudes 
give a $\sim$30\% contribution;
(iii) the $NN$ rescattering contribution is not well suppressed
($\sim$10\% contribution) and thus, considering the precision in question,
a contribution from multiple
rescatterings beyond the first-order rescattering would be nonnegligible.
On the other hand, 
our result obtained at the ELPH kinematics
is essentially free from these contributions (i)-(iii) that are currently
difficult to control with a high precision.
Another benefit of utilizing the ELPH kinematics is that 
the cross sections are fairly large near the QF peak,
making a precise measurement possible.
Indeed, our cross sections at the QF peak in Fig.~\ref{fig:eta-a}
are $\sim$20 times larger than those shown in 
Fig.~6 (bottom) of Ref.~\cite{mainz2004}.
One more advancement
is that 
we proposed to use the ratio, Eq.~(\ref{eq:Ratio}), to
cancel out the $\sim$5\% 
uncertainty inherent in any elementary 
$\gamma p\to \eta p$ amplitudes.
The advancements described above lead us to a
conclusion
that it will be possible to significantly improve the precision of 
the $\eta N$ scattering parameters using the ELPH data.

In conclusion, 
we have analyzed the $\gamma d\to\eta pn$ reaction 
at $E_\gamma=0.94$~GeV and $\theta_p=0^\circ$, and found
that, 
once $R_{\rm expt}$ data of 5\% error binned in 1 MeV width are given, 
${\rm Re}[a_{\eta N}]$ (${\rm Re}[r_{\eta N}]$)
can be determined 
at the precision of $\sim\pm 0.1$~fm ($\sim\pm 0.5$~fm),
which is significantly better than the currently estimated uncertainty. 
We emphasize that, for reliably extracting the $\eta N$
scattering parameters
from the data, it is prerequisite to
control all the relevant subprocess in
$\gamma d\to\eta pn$ with a sophisticated model like
the DCC model~\cite{knls13,knls16}.

\begin{acknowledgments}
We thank T. Sato for a useful discussion.
This work was supported in part by JSPS KAKENHI Grant
Nos.~JP25105010, JP25800149,
and JP26400287, 
and
Funda\c{c}\~ao de Amparo \`a Pesquisa do Estado de S\~ao Paulo-FAPESP,
Process No.~2016/15618-8.
\end{acknowledgments}

\end{document}